\documentclass[aps,pra,showpacs]{revtex4}

\usepackage{dcolumn}
\usepackage{graphicx}

\begin{document}

\title{Prony's method and the connected--moments expansion}
\author{Francisco M. Fern\'{a}ndez}

\affiliation{INIFTA (UNLP, CCT La Plata--CONICET), Blvd. 113 y 64 S/N, \\
Sucursal 4, Casilla de Correo 16, 1900 La Plata, Argentina}

\begin{abstract}
We show that Prony's method provides the full solution to the nonlinear
equations of the connected--moments expansion (CMX). Knowledge of all the
parameters in the CMX ansatz is useful for the analysis of the convergence
properties of the approach. Prony's method is also suitable for the
calculation of the correlation function for simple quantum--mechanical
models.
\end{abstract}


\maketitle

\section{Introduction}

Horn and Weinstein\cite{HW84} and Horn et al\cite{HKW85} proposed the $t$%
--expansion for the calculation of the ground--state energy of
quantum--mechanical models. It is the Taylor expansion about $t=0$ of a
monotonically decreasing function $E(t)$ that leads to the ground--state
energy when $t\rightarrow \infty $ provided that the chosen reference
function exhibits a nonzero overlap with the ground state. The coefficients
of such Maclaurin series are known as connected moments or cummulants.

Since the extrapolation of the $t$--expansion towards $t\rightarrow \infty $
by means of Pad\'{e} approximants\cite{HW84,HKW85} did not appear to produce
encouraging results, Cioslowski\cite{C87a} proposed an exponential series
and Stubbins\cite{S88} compared it with other extrapolation approaches. On
matching the $t$--expansion and the exponential series at origin one has to
solve a system of nonlinear equations. In order to bypass this problem
Cioslowski\cite{C87a} developed a systematic algorithm for obtaining just
the parameter related to the energy by removing all the other variables in
the nonlinear equations. The resulting approach is known as connected
moments expansion (CMX). From the properties of the Pad\'{e} approximants
Knowles\cite{K87} derived a compact an ellegant explicit expression for the
approximants to the energy in terms of the connected moments.

Since the CMX approximants may exhibit
singularities\cite{K87,MBM89,MPM91} other authors proposed
improved approaches like the alternate moments expansion
(AMX)\cite{MZM94, MZMMP94} and the generalized moments
expansion(GMX)\cite {MMFB05}. Although these methods may avoid the
singularities in the CMX approximants they do not seem to improve
the convergence properties of the sequences of approximants and
commonly their results are not more accurate than the CMX ones.

A recent analysis of the convergence properties of the CMX
required the calculation of all the variables in the CMX
ansatz\cite{AFR11a}. For that reason, Amore and
Fern\'{a}ndez\cite{AF11b} proposed a systematic method for the
full solution of the CMX nonlinear equations motivated by an
earlier discussion\cite{F09} of the connected--moments polynomial
approach\cite{B08}.

Nonlinear equations like the CMX ones are an old problem in
applied mathematics and were first solved by Prony\cite{P1795} and
later Weiss and McDonough\cite{WM63} proposed an alternative
approach based on Pad\'{e} approximants. In fact, Prony's method
is well known in numerical analysis\cite{H74} and has been widely
applied to a variety of problems in chemistry and
engineering\cite{FMT07,HP02}. Further inspection of the procedure
developed by Amore and Fern\'{a}ndez\cite{AF11b} has revealed that
it is closely related to Prony's method, although the main result
derived by the former authors does not appear in the discussions
of the latter method\cite {WM63,H74}.

The main purpose of this paper is to show the connection between
Prony's method and the procedure developed by Amore and
Fern\'{a}ndez. In Sec.~\ref {Sec:Prony} we show how to solve a
particular set of nonlinear equations by means of Prony's method
and derive the main result of Amore and Fern\'{a}ndez\cite{AF11b}.
In Sec.~\ref{Sec:Generating} we discuss the application of that
method to the extrapolation of the generating functions for the
moments and connected moments. In Sec.~\ref{Sec:examples} we test
the general analytical results on simple models and show that they
are useful for a discussion of the convergence properties of the
CMX as well as for the approximate calculation of the
autocorrelation function.

\section{Prony's method}

\label{Sec:Prony}

In what follows we discuss the solution of the system of $2N$ equations
\begin{equation}
F_{k}=\sum_{n=1}^{N}A_{n}b_{n}^{k+s},\;k=1,2,\ldots ,2N  \label{eq:main_eqs}
\end{equation}
for the $2N$ unknowns $A_{n}$ and $b_{n}$, where $F_{k}$ are known real
numbers and $s$ an integer.

Prony's method is based on the construction of the polynomial\cite{WM63,H74}
\begin{equation}
p(b)=\prod_{n=1}^{N}\left( b-b_{n}\right) =\sum_{j=0}^{N}p_{j}b^{j},\;p_{N}=1
\label{eq:poly}
\end{equation}
where the polynomial roots $b_{n}$, and thereby the polynomial coefficients $%
p_{j}$, are unknown. It follows from
\begin{equation}
\sum_{j=0}^{N}F_{i+j}p_{j}=\sum_{n=1}^{N}A_{n}b_{n}^{i+s}%
\sum_{j=0}^{N}p_{j}b_{n}^{j}=0,\;i=1,2,\ldots ,N
\end{equation}
that the coefficients $p_{j}$ are solutions to the linear system of
equations
\begin{equation}
\sum_{j=0}^{N-1}F_{i+j}p_{j}=-F_{i+N},\;i=1,2,\ldots ,N  \label{eq:linear_pj}
\end{equation}
If the matrix $\mathbf{F}$ with elements $F_{i+j}$, $i=1,2,\ldots ,N$, $%
j=0,1,\ldots ,N-1$ is nonsingular then the solution is unique. Once we have
the polynomial coefficients $p_{j}$ we obtain its roots $b_{n}$ and then
solve the resulting system of linear equations (\ref{eq:main_eqs}) (for
example for $k=1,2,\ldots ,N$) for the remaining unknowns $A_{n}$.

The technique just outlined is known since 1795\cite{P1795} and is commonly
called Prony's method\cite{WM63,H74}. Weiss and McDonough\cite{WM63}
proposed an alternative way of solving the nonlinear equations by means of a
partial fraction expansion of Pad\'{e} approximants, and in what follows we
describe a different strategy for obtaining the roots $b_{n}$ developed by
Amore and Fern\'{a}ndez\cite{AF11b}. The starting point of this procedure is
the homogeneous system of $N$ linear equations
\begin{equation}
\sum_{i=1}^{N}\left( F_{i+j}-bF_{i+j-1}\right) c_{i}=0,\;j=1,2,\ldots ,N
\label{eq:secular_eqs}
\end{equation}
with $N$ unknowns $c_{i}$. There will be nontrivial solutions ($c_{i}\neq 0$%
) only if the determinant vanishes:
\begin{equation}
\left| F_{i+j}-bF_{i+j-1}\right| _{i,j=1}^{N}=0  \label{eq:secular_det}
\end{equation}
Under such conditions we define
\begin{equation}
\gamma _{j}=\sum_{i=1}^{N}F_{i+j}c_{i}
\end{equation}
so that equations (\ref{eq:secular_eqs}) become $\gamma _{j}=b\gamma _{j-1}$
that leads to $\gamma _{j}=b^{j}\gamma _{0}$, $j=1,2,\ldots ,N$. Therefore,
it follows from
\begin{equation}
\sum_{j=0}^{N}\gamma
_{j}p_{j}=\sum_{i=1}^{N}c_{i}\sum_{j=0}^{N}F_{i+j}p_{j}=\gamma
_{0}\sum_{j=0}^{N}p_{j}b^{j}
\end{equation}
that the roots of the determinant (\ref{eq:secular_det}) are exactly the
roots of the polynomial $p(b)$ of Prony's method.

We have now at least three alternative procedures for solving the nonlinear
equations (\ref{eq:main_eqs}). First, the original Prony's method that
consists of solving the system of linear equations (\ref{eq:linear_pj}) for
the coefficients of the polynomial $p(b)$, then calculating its roots $b_{n}$%
, and finally solving the resulting system of linear equations (\ref
{eq:main_eqs}) for the remaining unknowns $A_{n}$. Second, the partial
fraction expansion of the Pad\'{e} approximants proposed by Weiss and
McDonough\cite{WM63}. Third, our approach that consists of obtaining the
polynomial $p(b)$ from the determinant~(\ref{eq:secular_det}). The choice of
either of them is probably a matter of taste. We find our technique quite
straightforward and it has revealed a most interesting connection between
the exponential expansion of the generating function for the moments and the
Rayleigh--Ritz method in the Krylov space\cite{AF11b}.

Prony's method was developed for fitting a series of exponential functions
to given numerical data\cite{P1795,WM63,H74,FMT07}. In what follows we apply
it to a completely different problem so that the name Prony's method refers
only to the strategy for solving the nonlinear equations (\ref{eq:main_eqs}).

\section{Generating functions for the moments and connected moments}

\label{Sec:Generating}

The general equations developed in the preceding section prove useful for
the analysis of the CMX\cite{C87a,K87}. We first consider the
moments--generating function\cite{HW84}
\begin{equation}
Z(t)=\left\langle \phi \right| e^{-t\hat{H}}\left| \phi \right\rangle
\label{eq:Z(t)}
\end{equation}
where $\hat{H}$ is the Hamiltonian operator of the system and $\left| \phi
\right\rangle $ is an arbitrary reference state. The coefficients of its
Maclaurin series
\begin{equation}
Z(t)=\sum_{j=0}^{\infty }\frac{(-t)^{j}}{j!}\mu _{j}  \label{eq:Z(t)_Taylor}
\end{equation}
are the moments $\;\mu _{j}=\left\langle \phi \right| \hat{H}^{j}\left| \phi
\right\rangle $.

For simplicity we assume that the spectrum of $\hat{H}$ is discrete
\begin{equation}
\hat{H}\left| \psi _{j}\right\rangle =E_{j}\left| \psi _{j}\right\rangle
,\;j=0,1,\ldots
\end{equation}
where $E_{0}\leq E_{1}\leq E_{2}\leq \ldots $, and without loss of
generality we choose the eigenvectors of $\hat{H}$ to be orthonormal: $%
\left\langle \psi _{i}\right| \left. \psi _{j}\right\rangle =\delta _{ij}$.
Under such conditions the generating function exhibits the exponential
behaviour
\begin{equation}
Z(t)=\sum_{j=0}^{\infty }\left| \left\langle \phi \right| \left. \psi
_{j}\right\rangle \right| ^{2}e^{-tE_{j}}  \label{eq:Z(t)_exp_exp}
\end{equation}
where the expansion coefficients $\left| \left\langle \phi \right| \left.
\psi _{j}\right\rangle \right| ^{2}$and the energies $E_{j}$ are unknown. We
can calculate them approximately by means of the method developed in the
preceding section and the ansatz
\begin{equation}
Z_{N}(t)=\sum_{j=0}^{N-1}A_{j}e^{-tW_{j}}  \label{eq:Z_N(t)}
\end{equation}
where $A_{j}$ and $W_{j}$ are the approximations to $\left| \left\langle
\phi \right| \left. \psi _{j}\right\rangle \right| ^{2}$and $E_{j}$,
respectively. If we require that the Maclaurin expansion of this ansatz
yields the first moments $\mu _{j}$ exactly we are left with the nonlinear
equations (\ref{eq:main_eqs}) with $F_{k}=\mu _{k-1}$, $s=-1$ and $%
W_{j}=b_{j-1}$.

It is has been proved that matching those expansions at origin is equivalent
to the application of the Rayleigh--Ritz variational method in the Krylov
space (RRK)\cite{AF11b}. Note that the determinant (\ref{eq:secular_det})
becomes the secular determinant of the RRK. The RRK solutions
\begin{equation}
\left| \varphi _{j}\right\rangle =\sum_{i=0}^{N-1}c_{ij}\left| \phi
_{i}\right\rangle ,\;\left| \phi _{i}\right\rangle =\hat{H}^{i}\left| \phi
\right\rangle   \label{eq:varphi_j}
\end{equation}
satisfy
\begin{equation}
\left\langle \phi _{k}\right| \hat{H}\left| \varphi _{j}\right\rangle
=W_{j}\left\langle \phi _{k}\right| \left. \varphi _{j}\right\rangle
\label{eq:RRK}
\end{equation}
and the application of Prony's method in the way just outlined yields $%
A_{j}=\left| \left\langle \phi \right| \left. \varphi _{j}\right\rangle
\right| ^{2}$.

The CMX is based on the generating function
\begin{equation}
E(t)=-\frac{Z^{\prime }(t)}{Z(t)}  \label{eq:E(t)}
\end{equation}
that is monotonically decreasing\cite{HW84} and the coefficients of its
Maclaurin series are the connected moments $I_{j}$:
\begin{equation}
E(t)=\sum_{j=0}^{\infty }\frac{(-t)^{j}}{j!}I_{j+1}  \label{eq:t_exp}
\end{equation}
The recurrence relation\cite{HW84}
\begin{eqnarray}
I_{1} &=&\mu _{1}  \nonumber \\
I_{j+1} &=&\mu _{j+1}-\sum_{i=0}^{j-1}\left(
\begin{array}{c}
j \\
i
\end{array}
\right) I_{i+1}\mu _{j-i},\;j=1,2,\ldots  \label{eq:I_j(muj)}
\end{eqnarray}
yields the connected moments (or cummulants) $I_{i}$ in terms of the moments
$\mu _{j}$.

The main interest in $E(t)$ is that it provides a size consistent approach
to the ground--state energy\cite{HW84}
\begin{equation}
\lim_{t\rightarrow \infty }E(t)=E_{0}
\end{equation}
when $\left\langle \phi \right| \left. \psi _{0}\right\rangle \neq 0$. In
order to carry out this extrapolation Cioslowski\cite{C87a} proposed the
exponential--series ansatz
\begin{equation}
E^{(N)}(t)=A_{0}+\sum_{j=1}^{N}A_{j}e^{-b_{j}t}  \label{eq:E^(N)(t)}
\end{equation}
where the unknown parameters $b_{j}$ are supposed to be real and positive
and $A_{0}$ is the approximation to $E_{0}$. Matching this expression with
the $t$--expansion (\ref{eq:t_exp}) leads to the set of equations
\begin{equation}
A_{0}=I_{1}-\sum_{n=1}^{N}A_{n}  \label{eq:A0_I1-An}
\end{equation}
and
\begin{equation}
I_{k+1}=\sum_{n=1}^{N}A_{n}b_{n}^{k},\;k=1,2,\ldots ,2N
\label{eq:I_j_A_j_b_j}
\end{equation}
We can solve the set of $2N$ equations (\ref{eq:I_j_A_j_b_j}) by means of
the method of the preceding section with $F_{k}=I_{k+1}$ and $s=0$. In this
case the exponential parameters $b_{j}$, $j=1,2,\ldots ,N$ are the roots of
the pseudo--secular determinant
\begin{equation}
\left| I_{i+j+1}-bI_{i+j}\right| _{i,j=1}^{N}=0  \label{eq:secular_I_ij}
\end{equation}
Once we have them we solve $N$ of the resulting $2N$ linear equations (\ref
{eq:I_j_A_j_b_j}) for the coefficients $A_{j}$, $j=1,2,\ldots ,N$, and then
we obtain $A_{0}$ from equation\ (\ref{eq:A0_I1-An}). Cioslowski\cite{C87a}
and Knowles\cite{K87} developed remarkable strategies for obtaining $A_{0}$
without calculating the other parameters explicitly. We have just shown that
the explicit calculation of all the parameters in the exponential--series
ansatz (\ref{eq:E^(N)(t)}) is quite straightforward.

\section{Illustrative examples}

\label{Sec:examples}

In order to test the equations developed in the preceding section
and show their usefulness in the analysis of the CMX we first
consider an exactly solvable model: the dimensionless harmonic
oscillator
\begin{equation}
\hat{H}=-\frac{d^{2}}{dx^{2}}+x^{2}  \label{eq:H_HO}
\end{equation}
As a trial function we choose one of the examples discussed in earlier papers%
\cite{AF11b,AFR11a}

\begin{equation}
\phi (x)=\left\langle x\right| \left. \phi \right\rangle =\left( x^{2}-\frac{%
1}{2}\right) e^{-2x^{2}/5}  \label{eq:phi(x)_HO}
\end{equation}
for which $\left| \left\langle \phi \right| \left. \psi
_{2}\right\rangle \right| >\left| \left\langle \phi \right| \left.
\psi _{j}\right\rangle \right| $, $j\neq 2$. (The normalization
factor is irrelevant to present purposes) One advantage of this
example is that it is not difficult to obtain the generating
function exactly

\begin{equation}
E(t)=\frac{121u^{3}+189199u^{2}+8180919u+6561}{\left( 81-u\right) \left(
121u^{2}+20198u+81\right) },\;u=e^{-4t}  \label{eq:E(t)_HO_exact}
\end{equation}
Note that $\lim_{t\rightarrow \infty }E(t)=E_{0}=1$ because $\left\langle
\phi \right| \left. \psi _{0}\right\rangle \neq 0$ in agreement with the
discussion in Sec.~\ref{Sec:Generating}.

\begin{figure}[h]
\begin{center}
\bigskip\bigskip\bigskip \includegraphics[width=9cm]{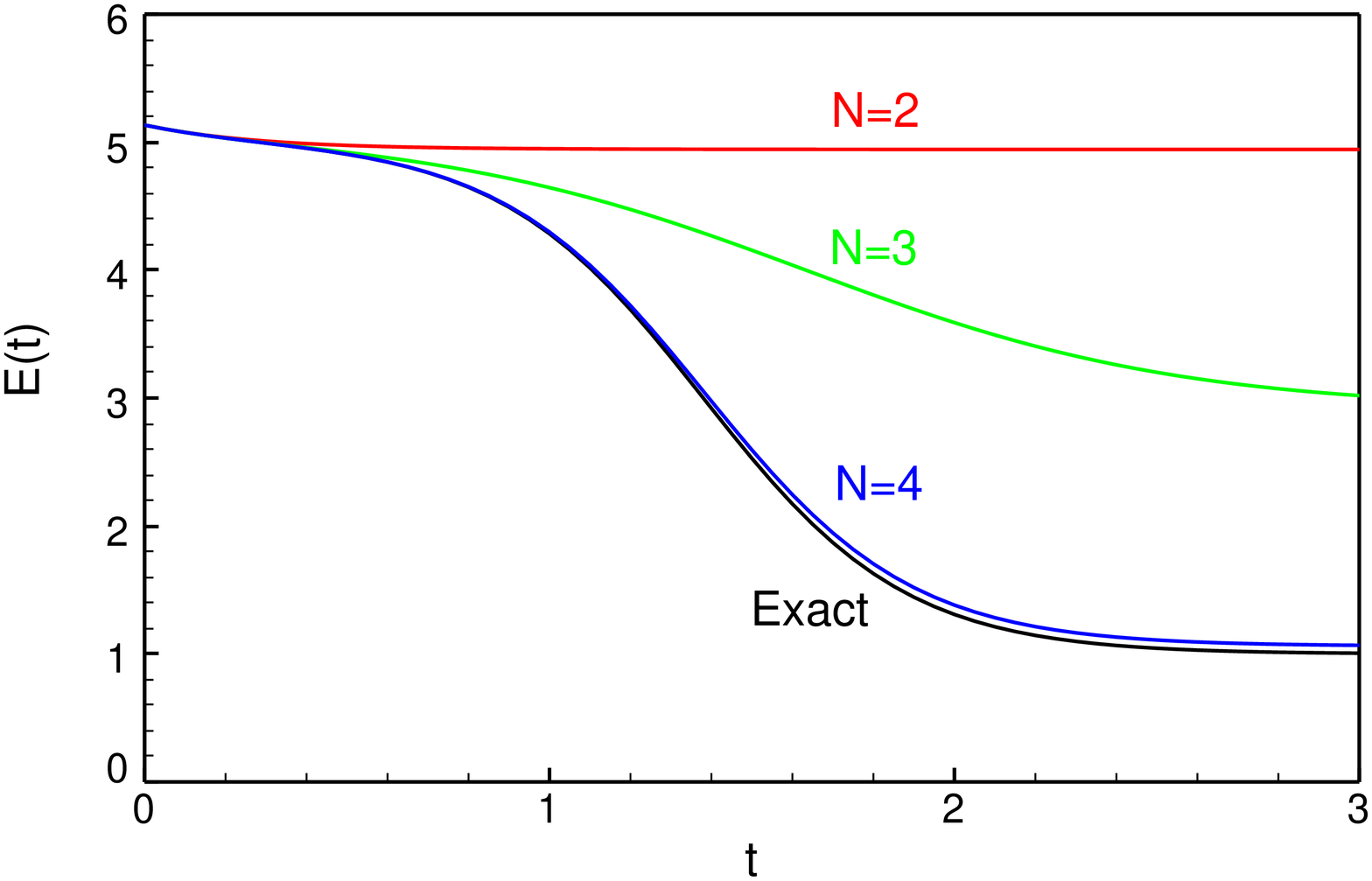}
\end{center}
\caption{Exact $E(t)$ and approximants $U^{(N)}(t)$ for the harmonic
oscillator }
\label{Fig:ETHO2A}
\end{figure}

\begin{figure}[h]
\begin{center}
\bigskip\bigskip\bigskip \includegraphics[width=9cm]{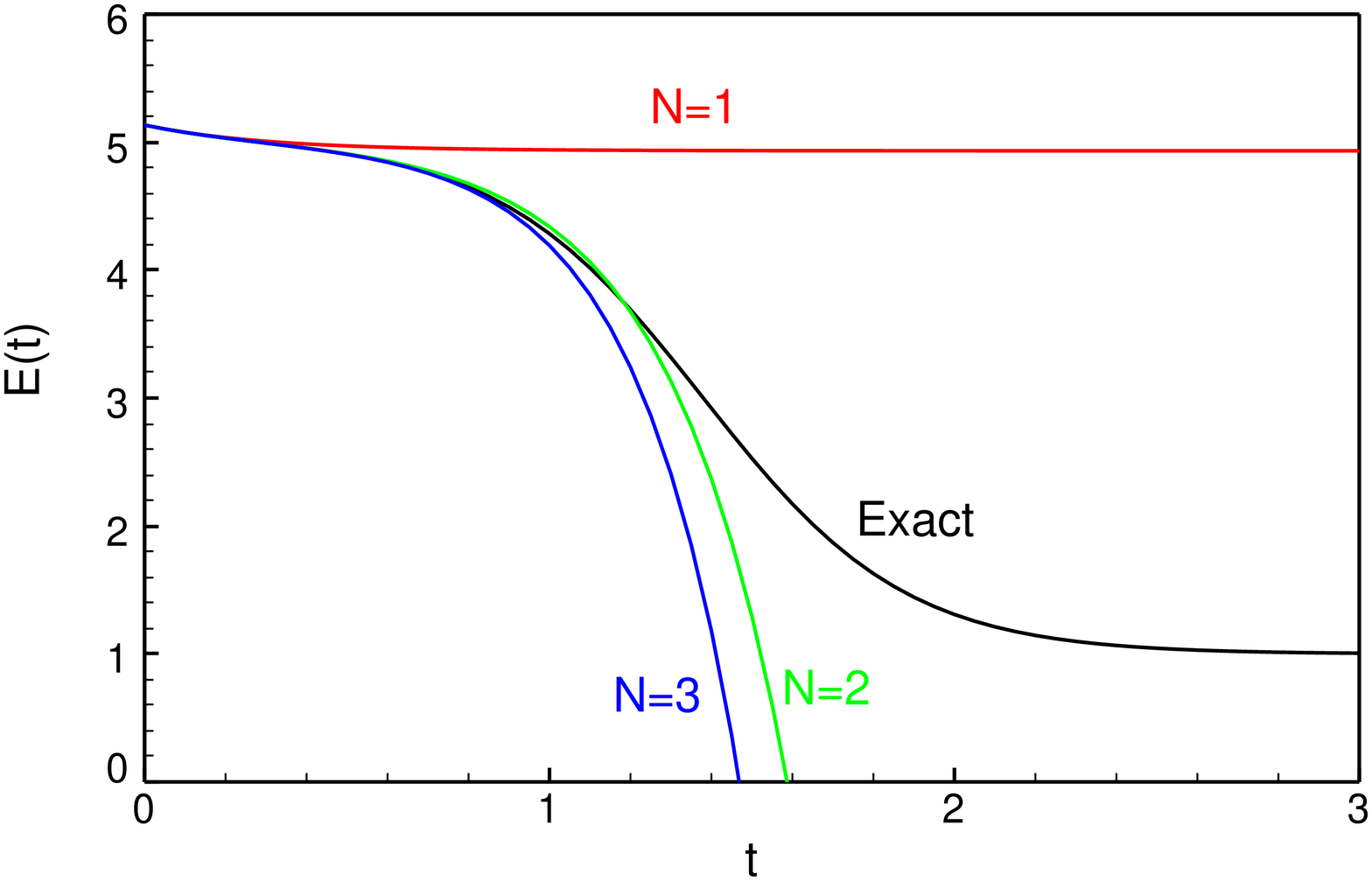}
\end{center}
\caption{Exact $E(t)$ and approximants $E^{(N)}(t)$ for the harmonic
oscillator }
\label{Fig:ETHO2B}
\end{figure}

By means of the general results of the preceding section we can
easily calculate the generating function $E(t)$ approximately in
two ways: as $U^{(N)}(t)=-Z_{N}^{\prime }(t)/Z_{N}(t)$ from
equation (\ref{eq:Z_N(t)}) and directly from equation
(\ref{eq:E^(N)(t)}). Fig~\ref{Fig:ETHO2A} shows that $U^{(N)}(t)$
approaches the exact generating function $E(t)$ for all $0\leq
t<\infty $ as $N$ increases. On the other hand, the approximate
expression $E^{(N)}(t)$ given by Eq.~(\ref{eq:E^(N)(t)}) is an
unsatisfactory approximation to the exact $E(t)$ as shown in
Fig.~\ref{Fig:ETHO2B} (except for sufficiently small values of
$t$). This situation does not appear to improve as $N$ increases.
It is found that $\lim_{t\rightarrow \infty }E^{(N)}(t)=-\infty $
for $N=2,3$ because in both cases there is one negative root
$b_{j}$ associated to a negative coefficient $A_{j}$. For example,
$A_{1}\approx -0.0170$, $A_{2}\approx 0.147$, $A_{3}\approx
0.000617 $, $b_{1}\approx -3.87$, $b_{2}\approx 4.04$, and
$b_{3}\approx 9.29 $ for $N=3$. Note that the curves in both
figures are directly comparable in the sense that the calculation
of either $Z_{N+1}(t)$ and $E^{(N)}(t)$ requires exactly the same
number of moments $\mu _{j}$ ($2N+1$). The behaviour of
$E^{(N)}(t)$ for larger $N$ may be different; for example, for
$N=4,5$ there are two negative roots and $\lim_{t\rightarrow
\infty }E^{(N)}(t)=+\infty $.

The anomalous behaviour of $E^{(N)}(t)$ is due to the poles of
$E(t)$ in the complex $t$--plane. Note that
Eq.~(\ref{eq:E(t)_HO_exact}) exhibits three real poles in the
$u$--plane and the closest to the origin is located at $u\approx
-0.0040$. It is reasonable to assume that the CMX ansatz
$E^{(N)}(t) $ should approach the Taylor series about $u=0$ for
$E(t)$ when the approach is successful. However, this expansion
does not converge for $u=1$ which is necessary for matching the
$t$--expansion and exponential series at $t=0$. This problem was
discussed by Amore et al\cite{AFR11a} by means of two--level
models and we see it here again for the harmonic oscillator. Those
authors have shown that the CMX approximants for the harmonic
oscillator with the trial function (\ref{eq:phi(x)_HO}) converge
towards the second--excited state $E_{2}=5$. From the approximate
expressions $E^{(N)}(t) $ with $N=1,2,3$ we obtain $A_{0}\approx
4.932,5.015,5.002$, respectively, in agreement with those results.
This example shows that the CMX approximants may converge to a
meaningful result even when the ansatz $E^{(N)}(t)$ (on which the
approach is based) is an unacceptable global approximation to
$E(t)$.

Knowles\cite{K87} argued that if the connected moments $I_{k}$ are
positive for all $k$ then the $b_{i}$ are real and positive.
Later, Massano et al\cite {MBM89} and Mancini et al\cite{MPM91}
suggested that Knowles' argument may not be valid and that it is
the Hadamard determinants constructed from the $I_{k}$ that should
be positive. The example just analysed supports the latter
conclusion because the first connected moments are all positive
($\approx 5.13$, $0.665$, $2.12$, $11.2$, $40.0$, $216$, $979$)
but one of the roots is negative as shown above. We realize that
the general formulas derived in Sec.~\ref{Sec:Prony} are most
useful for this kind of analysis. In fact, the calculation of the
roots $b_{j}$ by means of present formulas is almost as easy as
the calculation of the Hadamard determinants and the former
provide a much clearer indication of the suitability of the ansatz
$E^{(N)}(t)$ as an approach for $E(t)$.

By means of the example discussed above we do not want to convey
the wrong impression that the CMX is suitable for the calculation
of excited--state energies because it is quite difficult to choose
an appropriate trial function for this purpose (except when we can
exploit the symmetry of the problem to make the trial function
orthogonal to the ground--state\cite {AF09b}). The main interest
in that exactly solvable problem is to show that we may predict an
anomalous behaviour of the CMX by means of the roots $b_{j} $. If
they are not real and positive it is reasonable to suspect that
the trial function may not be suitable and that it is convenient
to choose another one. For example, if $\phi (x)=e^{-x^{2}}$ all
the roots $b_{j}$ are positive and the resulting $E^{(N)}(t)$ is a
remarkably good approximation to $E(t)$ for all $0\leq t<\infty $
even at the low orders $N=1,2$.

Finally, we explore the possibility of calculating of the
correlation function
\begin{equation}
C(t)=Z(it)=\left\langle \phi \right| e^{-it\hat{H}}\left| \phi \right\rangle
\label{eq:C(t)}
\end{equation}
where $\left| \phi \right\rangle =\left| \phi (0)\right\rangle $
is the initial state normalized to unity and $\left| \phi
(t)\right\rangle =e^{-it\hat{H}}\left| \phi (0)\right\rangle $ is
the state at time $t$. In particular we concentrate on the real
function $\left| C(t)\right| ^{2}$.

\begin{figure}[h]
\begin{center}
\bigskip\bigskip\bigskip
\par
\includegraphics[width=9cm]{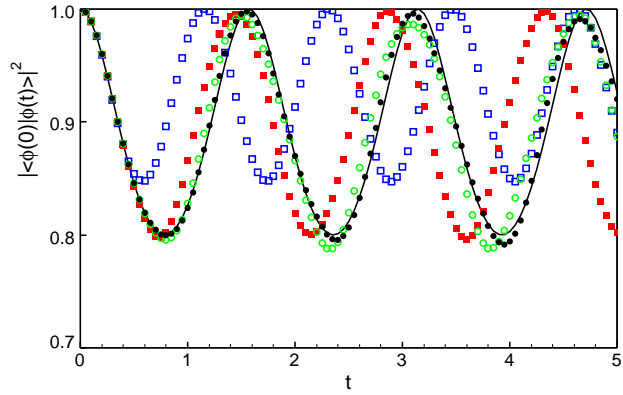}
\end{center}
\caption{Exact $\left|C(t)\right|^2$ for the harmonic oscillator (line) and
approximants $\left|Z_N(it)\right|^2$ with $N=2$ (squares,blue), $N=3$
(filled squares, red), $N=4$ (circles, green) and $N=5$ (filled circle,
black).}
\label{Fig:ZITHO}
\end{figure}

As a first example we consider the harmonic oscillator (\ref{eq:H_HO}) and
the trial function
\begin{equation}
\phi (x)=\left( \frac{2}{\pi }\right) ^{1/4}\mathbf{e}^{-x^{2}}
\label{eq:phi(0)_HO}
\end{equation}
The exact correlation function is

\begin{equation}
\left| C(t)\right| ^{2}=\frac{4\sqrt{2}}{\sqrt{41-9\cos {\left( 4t\right) }}}
\label{eq:C(t)^2_HO}
\end{equation}
Fig.~\ref{Fig:ZITHO} shows results for $N=2,3,4,5$. There is no
doubt that $\left| Z_{N}(it)\right| ^{2}$ converges towards
$\left| C(t)\right| ^{2}$ as $N$ increases. The approximation is
satisfactory within the first period but the errors accumulate as
$t$ increases. This fact is not surprising because the approximate
expression is constructed from the Maclaurin series for $Z(it)$.

\begin{figure}[h]
\begin{center}
\bigskip\bigskip\bigskip \includegraphics[width=9cm]{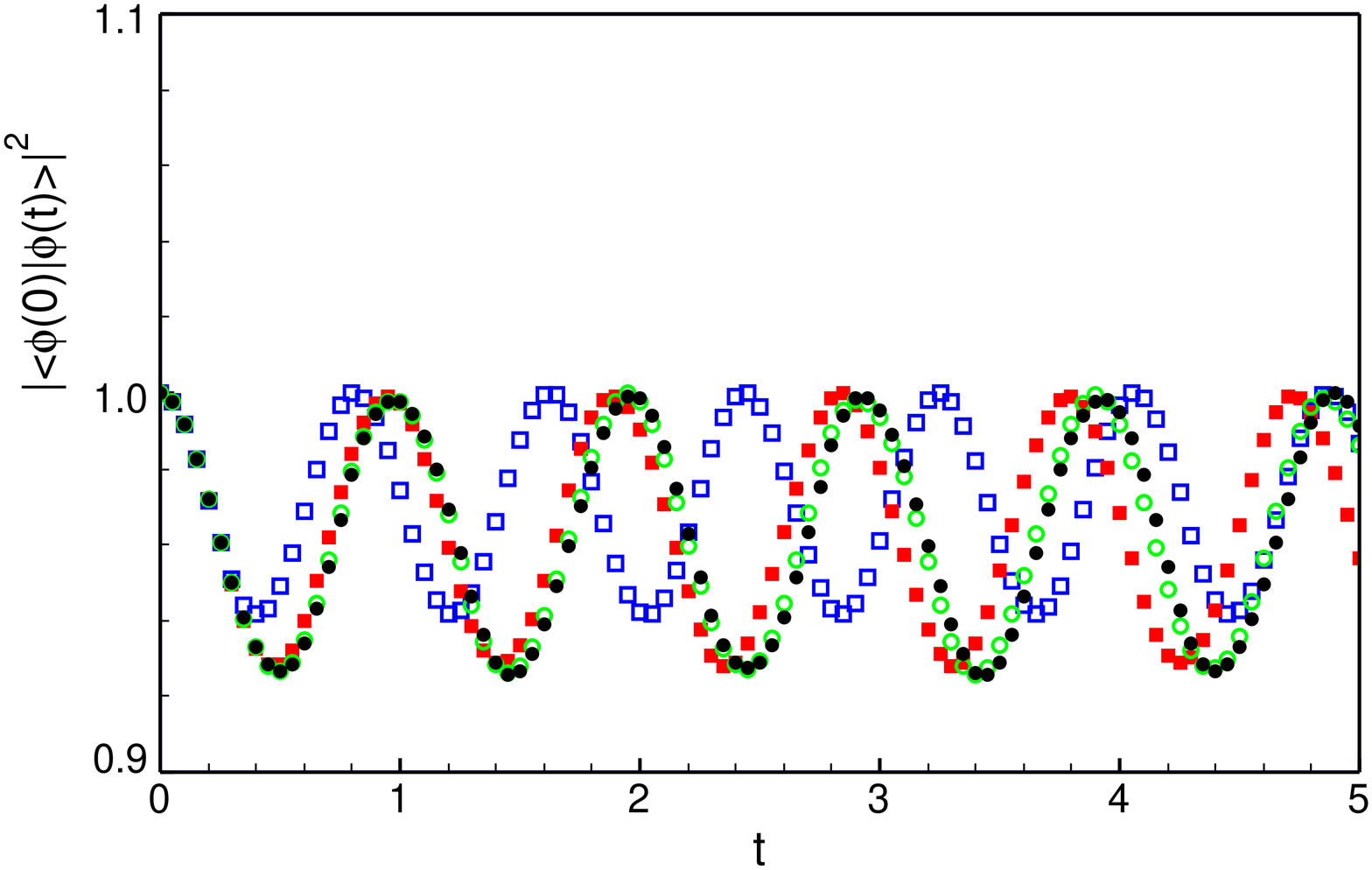}
\end{center}
\caption{Approximants $\left|Z_N(it)\right|^2$ for the anharmonic oscillator
(\ref{eq:H_AHO}) with $N=2$ (squares,blue), $N=3$ (filled squares, red), $%
N=4 $ (circles, green) and $N=5$ (filled circle, black).}
\label{Fig:ZITAHO}
\end{figure}

The second example is the anharmonic oscillator
\begin{equation}
\hat{H}=-\frac{d^{2}}{dx^{2}}+x^{4}  \label{eq:H_AHO}
\end{equation}
and the same trial function (\ref{eq:phi(0)_HO}).
Fig.~\ref{Fig:ZITAHO} shows that the approximate results for
$N=2,3,4,5$ follow the same trend as in the case of the harmonic
oscillator. At first sight it may be surprising that the
convergence rate for the anharmonic oscillator appears to be
greater than for the harmonic one. The reason is that the chosen
trial state exhibits a greater overlap with the ground state of
the former. Note that the magnitude of the coefficient $A_{0}$ of
$Z_{N}(t)$ tells us that $\left| \left\langle \phi \right. \left|
\psi _{0}\right\rangle \right| ^{2}$ is approximately 0.943 for
the harmonic oscillator and 0.981 for the anharmonic one.

\begin{figure}[h]
\begin{center}
\bigskip\bigskip\bigskip \includegraphics[width=9cm]{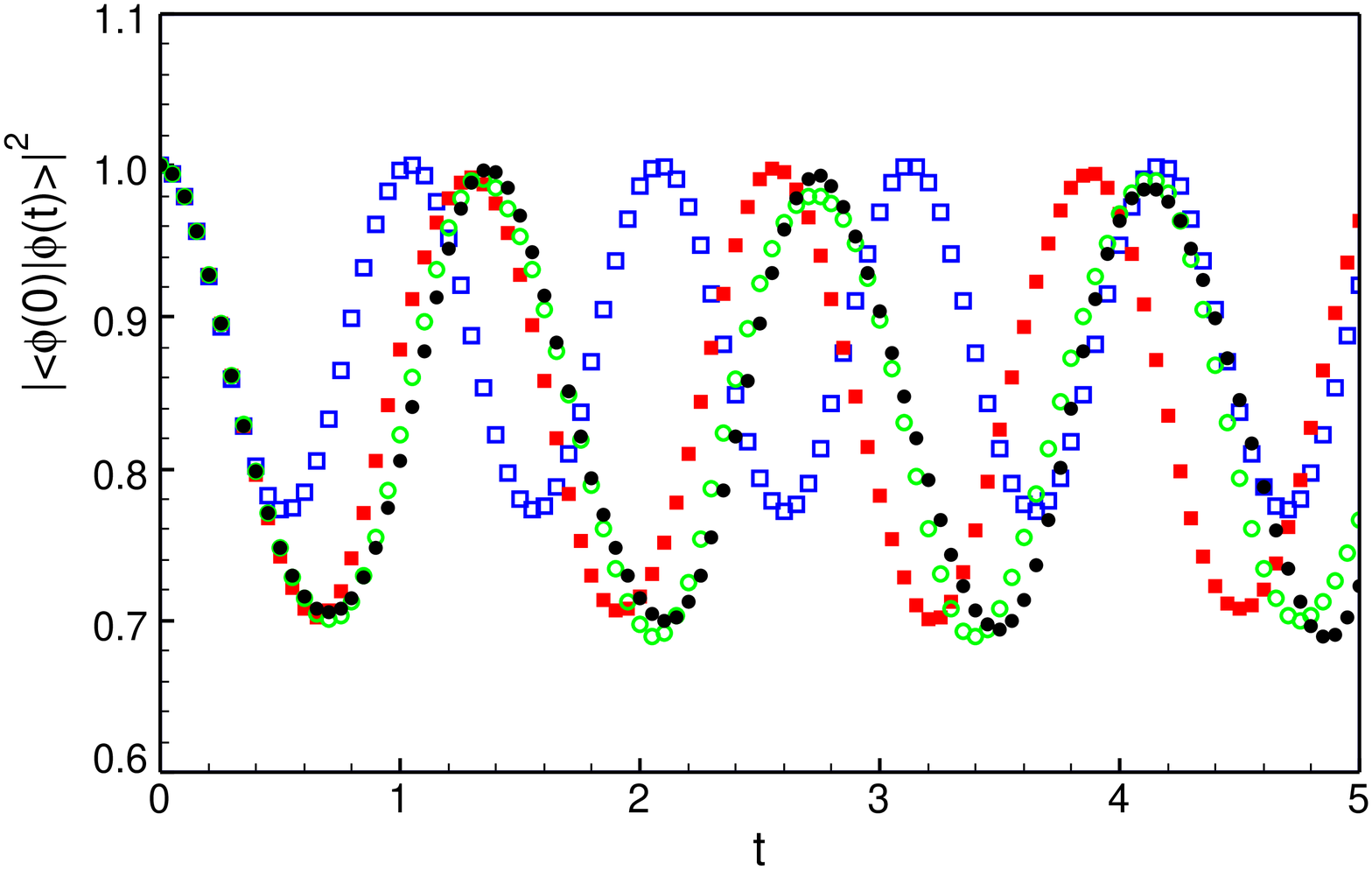}
\end{center}
\caption{Approximants $\left|Z_N(it)\right|^2$ for the anharmonic oscillator
(\ref{eq:H_PE}) with $N=2$ (squares,blue), $N=3$ (filled squares, red), $N=4$
(circles, green) and $N=5$ (filled circle, black).}
\label{Fig:ZITPE}
\end{figure}

As a two--dimensional example we consider the anharmonic oscillator
\begin{equation}
\hat{H}=-\frac{d^{2}}{dx^{2}}-\frac{d^{2}}{dy^{2}}+x^{2}+y^{2}+\lambda
x^{2}y^{2}  \label{eq:H_PE}
\end{equation}
where $\lambda >0$, and the trial function
\begin{equation}
\phi (x,y)=\left( \frac{2}{\pi }\right) ^{1/2}e^{-x^{2}-y^{2}}
\label{eq:phi(0)_PE}
\end{equation}
Fig.~\ref{Fig:ZITPE} shows results for $\lambda =0.5$ that look quite
similar to those for the one--dimensional models.

One does not expect that it may be possible to construct $C(t)$
from its Taylor expansion about $t=0$ in all the cases. The
problems discussed above are simple examples of single--well
oscillators. In the case of the double--well oscillator
$V(x)=(x^{2}-1)^{2}$ and the trial state localized in one of the
wells $\phi (x)=\left( \frac{2}{\pi }\right) ^{1/4}e^{-(x-1)^{2}}$
the approximants to $\left| C(t)\right| ^{2}$ do not appear to
converge although the results for $Z(t)$ look reasonable. If, on
the other hand, we locate the initial state on top of the barier
(for example, Eq.~(\ref{eq:phi(0)_HO})) then the results are as
accurate as those for the single wells.

\section{Conclusions}

\label{sec:conclusions}

The main goal of this paper is to show that\ Prony's method provides the
full solution of the CMX equations in a straightforward and simple way. We
have outlined the connection between Prony's method and the procedure
developed recently by Amore and Fern\'{a}ndez\cite{AF11b}. Although both
approaches are equivalent and Prony's one is known since long ago it seems
that the pseudo--secular determinants obtained by Amore and Fern\'{a}ndez
are not available elsewhere.

In spite of the fact that the harmonic oscillator with the trial function (%
\ref{eq:phi(x)_HO}) was chosen as an illustrative example in two earlier
papers\cite{AFR11a,AF11b} we think that present discussion based on the
exact generating function (\ref{eq:E(t)_HO_exact}) is clearer and provides
more information about the behaviour of the CMX ansatz $E^{(N)}(t)$. In
particular, we have shown that this simple problem clearly illustrates that
Knowles' conclusion about the sign of the connected moments\cite{K87} is not
valid as argued before by Massano et al\cite{MBM89} and Mancini et al\cite
{MPM91}.

It is clear that the roots $b_{i}$ provide a clear indication of the success
of the CMX and the method developed by Amore and Fern\'{a}ndez\cite{AF11b},
as well as Prony's method, make their calculation straightforward. Both
procedures may be applied to the calculation of the autocorrelation function
for simple quantum--mechanical models. If the approach converges then one
obtains a reasonable analytic expression for the autocorrelation function
within the first period of oscillation.

\end{document}